\documentstyle[12pt,epsfig,a4]{article} 
\newcommand{\vs}{\vspace{-0.25cm}}
\begin{document} 
\begin{center}
{\Large{\bf Chiral corrections to $\pi^-\gamma\to 3\pi$ processes  at low 
energies} }  
\bigskip

N. Kaiser \\
\medskip
{\small Physik-Department T39, Technische Universit\"{a}t M\"{u}nchen,
    D-85747 Garching, Germany}
\end{center}
\medskip
\begin{abstract}
We calculate in chiral perturbation theory the double-pion photoproduction 
processes $\pi^-\gamma \to \pi^- \pi^0\pi^0$ and $\pi^-\gamma\to \pi^+\pi^-\pi^-$ 
at low energies. At leading order these reactions are governed by the chiral 
pion-pion interaction. The next-to-leading order corrections arise from 
pion-loop diagrams and chiral-invariant counterterms involving  the low-energy 
constants $\bar\ell_1,\, \bar\ell_2,\, \bar\ell_3$ and $\bar\ell_4$. The
pertinent production amplitudes $A_1$ and $A_2$ depending on five kinematical
variables are given in closed analytical form. We find that the total 
cross section for neutral pion-pair production $\pi^-\gamma \to\pi^-\pi^0\pi^0$ 
gets enhanced in the region $\sqrt{s}< 7m_\pi$ by a factor $1.5 - 1.8$ by the 
next-to-leading order corrections. In contrast to this behavior the total cross 
section for charged pion-pair production  $\pi^-\gamma\to \pi^+ \pi^-\pi^-$ 
remains almost unchanged in the region $\sqrt{s}< 6m_\pi$ in comparison to its 
tree-level result. Although the dynamics of the pion-pair production reactions 
is much richer, this observed pattern can be understood from the different 
influence of the chiral corrections on the pion-pion final state interaction 
($\pi^+\pi^- \to \pi^0\pi^0$ versus $\pi^-\pi^- \to  \pi^-\pi^-$). We present 
also results for the complete set of two-pion invariant mass spectra. The 
predictions of chiral perturbation theory for the $\pi^-\gamma\to 3\pi$ 
processes can be tested by the COMPASS experiment which uses 
Primakoff scattering of high-energy pions in the Coulomb field of a heavy 
nucleus to extract cross sections for $\pi^-\gamma$ reactions with 
various final states.     
\end{abstract}

\bigskip

\section{Introduction and summary}
The pions $(\pi^+,\pi^0,\pi^-)$ are the Goldstone bosons of spontaneous chiral 
symmetry breaking in QCD: $SU(2)_L\times SU(2)_R \to SU(2)_V$. Their low-energy 
dynamics can therefore be calculated systematically (and accurately) with 
chiral perturbation theory in form of a loop-expansion based on an effective 
chiral Lagrangian. The accurate two-loop prediction \cite{cola} for the 
isospin-zero S-wave $\pi\pi$-scattering length $a_0^0 = (0.220 \pm 0.005)
m_\pi^{-1}$ has been confirmed in the E865 \cite{bnl} and NA48/2 \cite{batley} 
experiments by analyzing the $\pi^+\pi^-$ invariant mass distribution of the 
rare kaon decay mode $K^+ \to \pi^+\pi^-e^+\nu_e$. One particular implication 
of that good agreement between theory and experiment is that the quark 
condensate $\langle 0|\bar q q |0\rangle $ constitutes the dominant order 
parameter \cite{condensate} of spontaneous chiral symmetry breaking 
(considering the two-flavor sector of QCD). Likewise, the DIRAC experiment 
\cite{dirac} has been proposed to determine the difference of the isospin-zero 
and isospin-two S-wave $\pi\pi$-scattering lengths $a_0^0-a_0^2$ by measuring 
the life time ($\tau \simeq 3\,$fs) of pionium (i.e. $\pi^+\pi^-$ bound 
electromagnetically and decaying into $\pi^0\pi^0$). In the meantime the NA48/2
experiment \cite{cusp} has accumulated very high statistics for the charged 
kaon decay modes $K^\pm \to \pi^\pm \pi^0\pi^0$, which allowed to extract the 
value $a_0^0-a_0^2=(0.268 \pm 0.010)m_\pi^{-1}$ for the $\pi\pi$-scattering
length difference from the cusp effect in the $\pi^0\pi^0$ mass spectrum at the
 $\pi^+\pi^-$ threshold. This experimental result is again in very good 
agreement with the two-loop prediction $a_0^0-a_0^2 = (0.265 \pm 0.004)m_\pi^{-1}$ 
of chiral perturbation theory  \cite{cola}. For a discussion of isospin 
breaking corrections which have to be included in a meaningful comparison 
between theory and experiment, see ref.\cite{gasserisobr}. Clearly, these 
remarkable confirmations give confidence that chiral perturbation theory is
the correct framework to calculate reliably and accurately the strong
interaction dynamics of the pions at low energies.  

Electromagnetic processes offer further possibilities to probe the internal
structure of the pion. For example, pion Compton scattering $\pi^-\gamma\to
\pi^-\gamma$ at low energies allows one to extract the electric and magnetic 
polarizabilities ($\alpha_\pi$ and $\beta_\pi $) of the charged pion. Chiral 
perturbation theory at two-loop order gives for the dominant pion polarizability
difference the firm prediction $\alpha_\pi- \beta_\pi =(5.7\pm1.0)\cdot 10^{-4}\,
$fm$^3$ \cite{gasser}. It is however in conflict with the existing experimental 
results from Serpukhov $\alpha_\pi-\beta_\pi= (15.6\pm 7.8)\cdot 10^{-4}\,$fm$^3$ 
\cite{serpukov} and MAMI $\alpha_\pi-\beta_\pi=(11.6\pm 3.4)\cdot 10^{-4}\,$fm$^3$
\cite{mainz} which amount to values more than twice as large. Certainly, these
existing experimental determinations of $\alpha_\pi- \beta_\pi$ raise doubts
about their correctness since they violate the chiral low-energy theorem 
notably by a factor 2. The chiral low-energy theorem \cite{terentev} relates 
$\alpha_\pi- \beta_\pi=\alpha (\bar \ell_6-\bar \ell_5)/(24\pi^2 f_\pi^2m_\pi) 
+{\cal  O}(m_\pi)$ to the axial-vector-to-vector form factor ratio $h_A/h_V = 
0.443\pm 0.015 = (\bar \ell_6-\bar \ell_5)/6 +{\cal O}(m_\pi^2)$ measured in the 
PIBETA experiment \cite{frlez} via the radiative pion decay $\pi^+\to e^+\nu_e
\gamma$. The two-loop calculations of refs.\cite{gasser,buergi,geng}
assure that the ${\cal  O}(m_\pi)$ corrections to it are in fact small.

In that contradictory situation, it is promising that the ongoing COMPASS 
experiment \cite{compass} at CERN aims at remeasuring the pion polarizabilities, 
$\alpha_\pi$ and $\beta_\pi$, with high statistics using the Primakoff effect. 
The scattering of high-energy negative pions in the Coulomb field of a heavy 
nucleus (of charge $Z$) gives access to cross sections for $\pi^-\gamma$
reactions through the equivalent  photon method:
\begin{equation} {d \sigma \over ds\, dQ^2} = {Z^2 \alpha\over \pi(s-m_\pi^2)}
\, {Q^2-Q_{\rm min}^2 \over Q^4}\,\, \sigma_{\pi^- \gamma}(s) \,, 
\qquad Q_{\rm min} = {s-m_\pi^2 \over 2E_{\rm beam}}\,. \end{equation}    
Here, $Q$ denotes the momentum transferred by the virtual photon to the heavy
nucleus of charge $Z$, and one aims at isolating the Coulomb peak $Q\to 0$ from
the strong interaction background. The last factor $\sigma_{\pi^- \gamma}(s)$ is 
the total cross section for a $\pi^-\gamma$ reaction induced by real photons 
with $\sqrt{s}$ the corresponding $\pi^-\gamma$ center-of-mass energy. Note 
that eq.(1) applies in the same form to differential cross sections on both 
sides. The COMPASS experiment is set up to detect simultaneously various 
(multi-particle) hadronic final states which are produced in the Primakoff 
scattering process of high-energy pions. In addition to pion Compton 
scattering $\pi^-\gamma\to \pi^-\gamma$ (which is of primary interest for 
determining the pion polarizabilities $\alpha_\pi$ and $\beta_\pi$) the reaction 
$\pi^-\gamma\to \pi^-\pi^0$ serves as a test of the QCD chiral anomaly 
(i.e. the anomalous $VAAA$ rectangle quark diagram) by measuring the $\gamma 
3\pi$ coupling constant $F_{\gamma 3\pi}=e/(4\pi^2 f_\pi^3) = 9.72\,$GeV$^{-3}$. For the 
two-body process $\pi^-\gamma\to \pi^-\pi^0$  the one-loop \cite{bijnens,picross} 
and two-loop corrections \cite{hannah} of  chiral perturbation theory as well 
as QED radiative corrections \cite{ametller} have been worked out. Thus an
accurate theoretical framework is available to analyze the upcoming data.    
The $\pi^- \gamma$ reaction with three charged pions in the final state is used 
by the COMPASS collaboration in the energy range 1\,GeV$\,< \sqrt{s}<\,
$2.5\,GeV to study the spectroscopy of non-strange meson resonances 
\cite{boris} ($a_1(1260)$, $a_2(1320)$, $\pi_2(1670)$, $\pi(1800)$, $a_4(2040)$ 
etc.) and to search for so-called exotic  meson resonances \cite{exotic}
(e.g. $\pi_1(1600)$) with quantum  numbers different from simple (constituent)
quark-antiquark bound states. The statistics of the COMPASS experiment is 
actually so high that the event rates with three pions in the final state can 
even be continued downward to the threshold. The cross sections (and other
more exclusive observables) of the $\pi^- \gamma \to 3\pi$ reactions in the 
low-energy region $\sqrt{s}<1\,$GeV offer new possibilities to test the strong 
interaction dynamics of the pions as predicted by chiral perturbation theory. 
The total cross sections for the processes $\pi^-\gamma  \to \pi^- \pi^0 \pi^0$ 
and $\pi^-\gamma  \to \pi^+ \pi^- \pi^-$ at tree-level have been calculated 
recently in ref.\,\cite{picross} and it was found that the cross section for the 
charged channel ($\pi^+ \pi^- \pi^-$) comes out about a factor of 2.5 larger than 
the one for the neutral channel ($\pi^- \pi^0 \pi^0$). In both cases it is the 
chiral pion-pion interaction (together with the electromagnetic pion-photon 
coupling) which governs these reactions at leading order. A preliminary analysis 
\cite{privat} of the COMPASS data for $\pi^-\gamma  \to \pi^+ \pi^- \pi^-$ in 
the  low-energy range  $0.5\,{\rm  GeV} <  \sqrt{s}< 0.8\, {\rm }$GeV leads to 
a total cross section which seems to be in agreement with the tree-level result 
\cite{picross} of chiral perturbation theory. An arguable aspect of that analysis
is the absolute normalization which has to known well in order to convert count 
rates of events into a cross section. However, before any conclusions about an 
agreement between theory and experiment can be drawn the predictions of chiral 
perturbation theory need to be sharpened further by including higher orders in 
the small momentum expansion. 

This is precisely the purpose of the present paper. We evaluate for the 
double-pion photoproduction processes $\pi^-\gamma  \to \pi^- \pi^0 \pi^0$ and
$\pi^-\gamma  \to \pi^+ \pi^- \pi^-$ the next-to-leading order corrections as
they arise from pion-loop diagrams and chiral-invariant counterterms
(proportional to the low-energy constants $\ell_j$). Our paper is 
organized as follows. In section 2, we treat first the somewhat simpler case
of neutral pion-pair production. We give for individual diagrams the
analytical expressions for the pertinent production amplitudes, $A_1$ and $A_2$,
which depend on five independent Lorentz-invariant kinematical variables.
In this detailed exposition one can check the exact cancellation of ultraviolet 
divergences between pion-loops and counterterms. Combined with the tree-level 
amplitudes, $A_1$ and $A_2$ are employed to calculate the total cross section and 
the two-pion mass spectra for $\pi^-\gamma  \to \pi^- \pi^0 \pi^0$. We find that 
the total cross sections (as well as the two-pion mass spectra) get enhanced
by a factor $1.5 - 1.8$ after including the next-to-leading order chiral 
corrections. Although the dynamics of the whole process is much richer this
enhancement can be understood (in an approximate way) from the $\pi^+\pi^-
\to \pi^0\pi^0$  final state interaction. In section 3, the same analytical 
calculation is performed for charged pion-pair production $\pi^-\gamma\to\pi^+ 
\pi^- \pi^-$. In that case many more diagrams do contribute since the photon 
can now couple to all three outgoing (charged) pions. For that reason we 
restrict ourselves to specifying the finite parts of the pion-loop diagrams 
and the total counterterm contribution (rewritten in terms of the low-energy 
constants $\bar \ell_j$ which subsume the chiral logarithm $\ln(m_\pi/\lambda)$
generated by pion-loops). Interestingly, we find that the total cross section for 
$\pi^-\gamma \to\pi^+\pi^- \pi^-$ in the low-energy region $3m_\pi <\sqrt{s}<6 
m_\pi$ remains almost unchanged by the inclusion of the next-to-leading order 
corrections. Although the dynamics of the process is much richer the (known) 
weak influence of chiral corrections on the isospin-two $\pi^-\pi^- \to\pi^- 
\pi^-$ scattering length $a_0^2$ \cite{cola} can provide an explanation for 
this feature. In this case the $\pi^+\pi^-$ mass spectrum reveals better certain
dynamical details which get averaged out in the total cross section. We 
estimate also the uncertainties of the observables for the $\pi^-\gamma \to 3\pi$ 
reactions which are induced by the present errorbars of the low-energy constants 
$\bar\ell_j$, and  find about $\pm 5\%$. 

In summary, we have calculated the processes $\pi^-\gamma  \to \pi^- \pi^0\pi^0$
and $\pi^-\gamma  \to \pi^+ \pi^- \pi^-$ at next-to-leading order in chiral 
perturbation theory. The predictions for the total cross sections and other 
more exclusive observables can be tested soon by the COMPASS experiment at CERN.
 
\section{Neutral pion-pair production}
In this section we treat the neutral pion-pair production process: $\pi^-(p_1)
+\gamma(k,\epsilon\,) \to\pi^-(p_2)+\pi^0(q_1)+\pi^0(q_2)$. We choose for
the (transversal) real photon the Coulomb-gauge in the center-of-mass frame,
which entails the conditions $\epsilon \cdot p_1 =\epsilon \cdot k= 0$. These
conditions imply that all diagrams for which the photon couples to the 
in-coming pion $\pi^-(p_1)$ vanish identically. Furthermore, it is advantageous
to parametrize the special-unitary matrix-field $U$ in the chiral Lagrangian 
${\cal L}_{\pi\pi}$ through an interpolating pion-field $\vec \pi$ in the form 
$U =\sqrt{1-\vec  \pi^{\,2} /f_\pi^2} + i \vec \tau \cdot \vec \pi/f_\pi$. It has 
the consequence that no $\gamma 4\pi$ and $\gamma 6\pi$ contact vertices exist 
at leading order. Under these assumptions one is left with one single 
non-vanishing tree diagram shown in Fig.\,1. Let us recall the expression for 
the total cross section for $\pi^-\gamma \to \pi^-\pi^0 \pi^0$ at tree-level:    
\begin{eqnarray}\sigma_{\rm tot}(s)&=&{\alpha\over 32\pi^2 f_\pi^4 (s-m_\pi^2)^3} 
\int\limits_{2m_\pi \sqrt{s}}^{s+m_\pi^2-4m^2_{\pi^0}}\!\!dw\,\sqrt{s-w+m_\pi^2-4m_{\pi^0}^2
\over s-w  +m_\pi^2}\nonumber \\ && \times\,(s-w+m_\pi^2-m_{\pi^0}^2)^2 \Bigg\{w 
\ln{w+\sqrt{w^2 -4m_\pi^2 s}\over 2m_\pi  \sqrt{s}}-\sqrt{w^2-4m_\pi^2s}\Bigg\}\,.  
\end{eqnarray}
Here, $s=(p_1+k)^2$ denotes the squared center-of-mass energy, $\alpha=e^2/
4\pi= 1/137.036$ is the fine structure constant, and $f_\pi=92.4\,$MeV 
stands for the pion decay constant. In comparison to eq.(16) in ref.\cite{picross}
we have included (some) isospin breaking effects by distinguishing the mass of 
the  charged pion $m_\pi=139.570\,$MeV from the mass of the neutral pion 
$m_{\pi^0}=134.977\,$MeV. Since it turns out that these effects are very small 
(see Fig.\,7) we will perform the whole calculation in the limit of isospin 
symmetry. In order to present the next-to-leading order corrections from chiral
loops and counterterms one has to start from the general form of the T-matrix, 
which reads (in Coulomb-gauge): 
\begin{equation} T = {2e \over f_\pi^2} \Big[ \vec \epsilon \cdot \vec q_1\,  A_1 
+ \vec \epsilon \cdot \vec q_2 \,  A_2 \Big] \,.\end{equation}
In this decomposition $A_1$ and $A_2$ are two dimensionless production
amplitudes which depend on  $s=(p_1+k)^2$ and on four other independent 
(Lorentz-invariant) Mandelstam  variables:
\begin{equation} s_1=(p_2+q_1)^2\,,  \quad s_2=(p_2+q_2)^2\,, \quad  t_1=(q_1-k)^2
\,,  \quad t_2=(q_2-k)^2\,. \end{equation}
This set is very convenient for describing the permutation of the two identical 
neutral pions in the final state via $(s_1\leftrightarrow s_2,\, t_1\leftrightarrow
t_2)$. For most diagrammatic contributions $A_1=A_2$, but there are a few 
exceptions which require their separate listing and specification.

\medskip
\begin{figure}
\begin{center}
\includegraphics[scale=1.,clip]{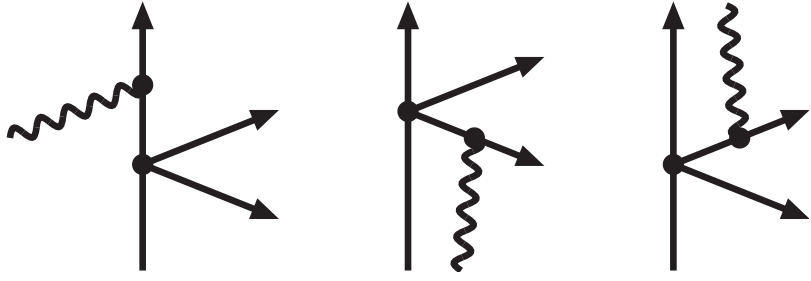}
\end{center}
\vspace{-.5cm}
\caption{Tree-level diagrams for $\pi^-\gamma  \to \pi^- \pi^0\pi^0$ and $ \pi^+ 
\pi^- \pi^-$. Arrows indicate out-going pions. Only the left diagram contributes
to $2\pi^0$-production.}
\end{figure}

With the form of the T-matrix defined in eq.(3) and the kinematical variables
introduced in eq.(4) the tree-level amplitudes read:
\begin{equation}A_1^{(\rm tree)} = A_2^{(\rm tree)} = {2m_\pi^2+s-s_1-s_2  \over 3m_\pi^2-s
-t_1-t_2} \,. \end{equation} 
Note that $T^{(\rm tree)}$ is written with physical parameters ($f_\pi^2$, 
$m_\pi^2$) and not with leading order parameters as given by the chiral 
Lagrangian. It requires an extra renormalization contribution (see eq.(26)) to 
properly account for this difference.      

\subsection{Diagrammatic calculation}
In this subsection we present analytical expressions for the amplitudes $A_1$
and $A_2$ as they arise from the next-to-leading order chiral loops and
counterterms. We go step by step through the whole set of contributing
diagrams. The one-loop diagrams (I), (II), (III) shown in Fig.\,2 include an 
additional $\pi\pi$-interaction but leave the pion-photon coupling (of the tree
diagram) unchanged. One finds from diagram (I): 
\begin{eqnarray}A_1^{(\rm I)}=A_2^{(\rm I)} &=&{1\over(4\pi f_\pi)^2}{2m_\pi^2+s-s_1-s_2 
 \over 3m_\pi^2-s-t_1-t_2}\bigg\{ \bigg(\xi + \ln{m_\pi \over \lambda} \bigg)
(s_1+s_2+t_1+t_2-11 m_\pi^2)  \nonumber \\ && + (s_1+s_2+t_1+t_2-7m_\pi^2)\bigg[ 
J(3m_\pi^2+s-s_1-s_2) -{1\over 2} \bigg] \bigg\} \,,\end{eqnarray} 
with the abbreviation 
\begin{equation} \xi = \lambda^{d-4}\bigg\{{1\over d-4} +{1\over  2}(\gamma_E-1-
\ln 4\pi)\bigg\}   \,, \end{equation} 
for standard ultraviolet divergence in dimensional regularization. Note that
$\xi$ is always accompanied by the chiral logarithm $\ln(m_\pi/\lambda)$. The 
complex-valued loop function $J(s)$ has the form:  
\begin{equation} J(s) = \sqrt{{s-4m_\pi^2\over s}}\,\Bigg[\ln {\sqrt{|s-4m_\pi^2|}
+ \sqrt{|s|} \over 2m_\pi} - {i\pi\over 2}\theta(s-4m_\pi^2) \Bigg] \,,\qquad 
s<0\,\,\,\,\,\,{\rm or}  \quad s>4m_\pi^2\,,  \end{equation} 
\begin{equation} J(s) = \sqrt{{4m_\pi^2-s \over s}} \, \arcsin{\sqrt{s} \over 
2m_\pi}  \,,\qquad 0<s<4m_\pi^2\,. \end{equation} 
Similarly, one gets from diagram (II):
\begin{eqnarray}A_1^{(\rm II)}=A_2^{(\rm II)} &=&{1\over 3(4\pi f_\pi)^2}{1\over 3m_\pi^2
-s-t_1-t_2}\bigg\{\bigg(\xi + \ln{m_\pi \over \lambda}\bigg)\Big[(s-s_2+t_1)(s-
s_1+s_2+2t_2)\nonumber \\ && +(5s_2+5t_2-4s_1-4t_1)m_\pi^2-13m_\pi^4\Big]+\Big[(s-
s_2+t_1)(s-s_1+s_2+2t_2)\nonumber \\ &&+(3s_2+t_2-2s-2s_1-4t_1)m_\pi^2+m_\pi^4\Big]
J(2m_\pi^2+s_2-s-t_1) +{1\over 6}(s-s_2+t_1)\nonumber \\ && \times (5s_1-5s-2s_2
-7t_2)+{m_\pi^2\over 6}(14s_1+11t_1-3s-4s_2-7t_2-7m_\pi^2)\bigg\}\,,\end{eqnarray} 
and the contribution from diagram (III) follows by interchanging the two
neutral pions:  
\begin{equation}A_1^{(\rm III)} = A_2^{(\rm III)} =A_1^{(\rm    II)}\Big|(s_1\leftrightarrow 
s_2,\, t_1\leftrightarrow t_2) \,.   \end{equation}    

\begin{figure}
\begin{center}
\includegraphics[scale=1.,clip]{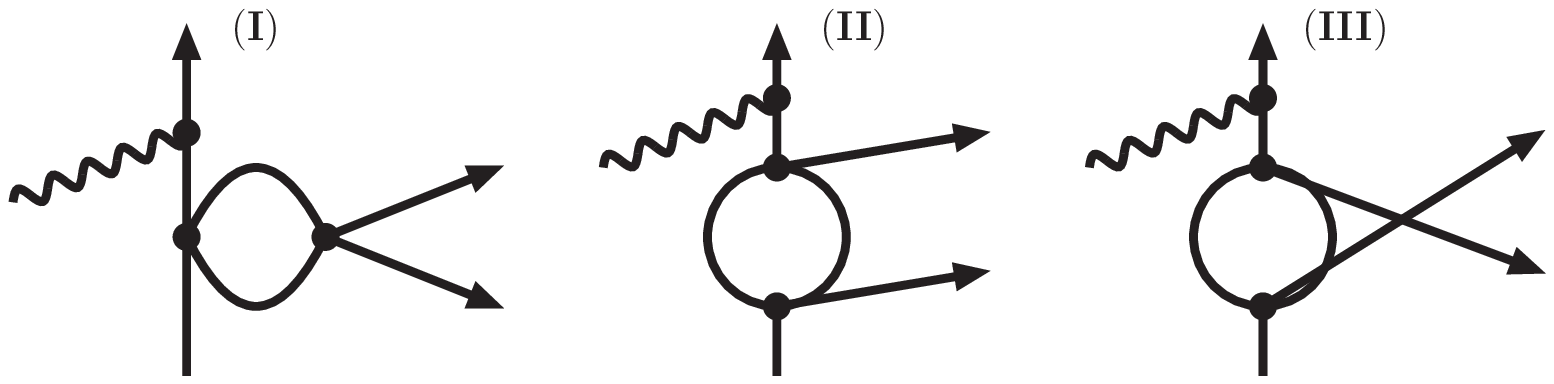}
\end{center}
\vspace{-.5cm}
\caption{One-pion loop diagrams for $\pi^-\gamma  \to \pi^- \pi^0\pi^0$. Their 
combinatoric factor is 1/2.}
\end{figure}

\begin{figure}
\begin{center}
\includegraphics[scale=1.,clip]{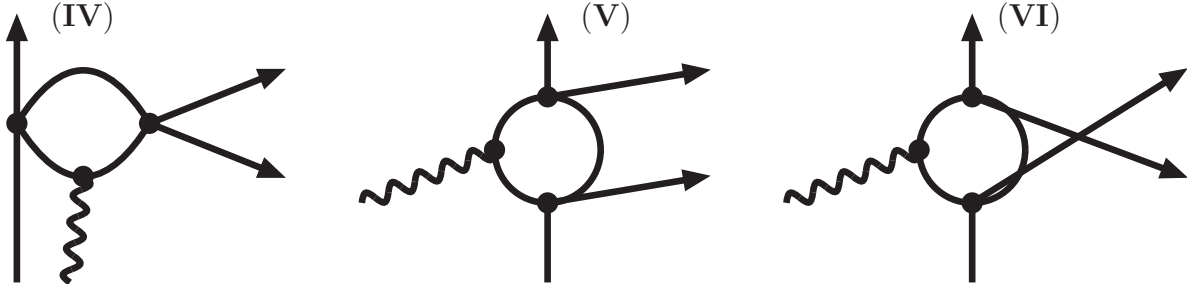}
\end{center}
\vspace{-.5cm}
\caption{One-pion loop diagrams for $\pi^-\gamma  \to \pi^- \pi^0\pi^0$. Their 
combinatoric factor is 1.}
\end{figure}
The one-loop diagrams (IV), (V), (VI) shown in Fig.\,3 include also an 
additional $\pi\pi$-interaction but the photon couples now to a charged pion 
inside the loop. One finds from diagram (IV):
\begin{eqnarray}A_1^{(\rm IV)}=A_2^{(\rm IV)} &=&{2m_\pi^2+s-s_1-s_2 \over (4\pi f_\pi)^2}
\Bigg\{\xi + \ln{m_\pi \over \lambda}- {1\over 2}+ J(3m_\pi^2+s-s_1-s_2) \nonumber 
\\ &&  +{m_\pi^2-s\over  2m_\pi^2-t_1 -t_2} +{2(s-m_\pi^2) \over (2m_\pi^2-t_1
-t_2)^2} \bigg\{(s_1+s_2-s-m_\pi^2-t_1-t_2) \nonumber \\ &&\times \Big[J(m_\pi^2+s
-s_1-s_2+t_1+t_2) -J(3m_\pi^2+s-s_1-s_2)  \Big]\nonumber \\ &&+ 2m_\pi^2 \Big[ 
G(m_\pi^2+s-s_1-s_2+t_1+t_2)-G(3m_\pi^2+s-s_1-s_2) \Big] \bigg\} \Bigg\}\,, 
\end{eqnarray}  
where the complex-valued loop function $G(s)$ has the form:
\begin{equation} G(s) = \Bigg[ \ln {\sqrt{|s-4m_\pi^2|}+ \sqrt{|s|} \over 
2m_\pi} - {i\pi\over 2}\theta(s-4m_\pi^2) \Bigg]^2 \,,\qquad 
s<0\,\,\,\,\,\,{\rm or}  \quad s>4m_\pi^2\,,  \end{equation} 
\begin{equation} G(s) = -\arcsin^2{\sqrt{s} \over 2m_\pi}  \,,\qquad  0<s<4m_\pi^2
\,. \end{equation}
Considerably more lengthy are the expressions which one obtains from evaluating 
diagram (V):
\begin{eqnarray}A_1^{(\rm V)} &=&{1 \over 3(4\pi f_\pi)^2} \Bigg\{\bigg(\xi + 
\ln{m_\pi \over \lambda}\bigg) (6m_\pi^2+2s-2s_1+t_2)+{1\over  6}(13s_1-13s-5t_2)
\nonumber \\ &&+{(s_1+8m_\pi^2)(m_\pi^2-t_2)\over 2(s+t_2-2m_\pi^2)}+6 m_\pi^2\Big[
G(2m_\pi^2+s_1-s-t_2)-G(s_1)\Big]\nonumber \\ && \times {  s(s+t_2-s_1)+(s_1-2s)
m_\pi^2 \over(s+t_2-2m_\pi^2)^2} +\bigg\{2s_1+4m_\pi^2+{s_1(s_1+8m_\pi^2)(t_2-m_\pi^2)
\over (s+t_2-2m_\pi^2)^2} \nonumber \\ && +{s_1(m_\pi^2-2t_2-2s_1)+4m_\pi^2(m_\pi^2-
t_2)\over s+t_2-2m_\pi^2}\bigg\}\Big[J(s_1)-J(2m_\pi^2+s_1-s-t_2)\Big] \nonumber 
\\ && + \bigg[2s-2s_1+t_2+{4m_\pi^2(t_2-m_\pi^2)\over  s+t_2-2m_\pi^2}\bigg]
J(2m_\pi^2+s_1-s-t_2)\Bigg\} \,, \end{eqnarray} 
\begin{eqnarray}A_2^{(\rm V)} &=&{1 \over 3(4\pi f_\pi)^2} \Bigg\{\bigg(\xi + 
\ln{m_\pi \over \lambda}\bigg)(s+s_1-s_2+t_1-3m_\pi^2)+{1\over  6}(5s_2+t_1-5s-8s_1)
\nonumber \\ && +{s_1(2t_2-3t_1)-t_1  t_2+m_\pi^2(s_1+2t_2-8t_1+7m_\pi^2)\over 2
(s+t_2-2m_\pi^2)}+{(s_1+8m_\pi^2)(t_1-m_\pi^2)(t_2-m_\pi^2)\over (s+t_2-2m_\pi^2)^2}
\nonumber \\ && +6m_\pi^2\Big[G(2m_\pi^2+s_1-s-t_2)-G(s_1)\Big]\bigg\{{s_1-t_1
\over  s+t_2-2m_\pi^2}+{2s_1(t_1-m_\pi^2)(m_\pi^2-t_2)\over (s+t_2-2m_\pi^2)^3} 
\nonumber \\ &&  +{s_1(2t_1-t_2)+t_1 t_2+m_\pi^2(m_\pi^2-s_1-2t_1)\over(s+t_2-
2m_\pi^2)^2}\bigg\}+\bigg[s-m_\pi^2+s_1-s_2+t_1\nonumber \\ && +{2(3t_1-t_2-2m_\pi^2)
\over s+t_2- 2m_\pi^2}+{8(t_1-m_\pi^2)(m_\pi^2-t_2)\over(s+t_2-2m_\pi^2)^2}\bigg]
J(2m_\pi^2+s_1-s-t_2)\nonumber \\ &&  +\bigg\{m_\pi^2-s_1+{s_1(3s_1-s_2-2t_1+t_2)
+m_\pi^2(3m_\pi^2+4s_2-4t_1-t_2) \over s+t_2-2m_\pi^2}\nonumber \\ && +{s_1^2(3t_1
-2t_2-m_\pi^2)+s_1[2t_1 t_2+m_\pi^2(7t_1-3t_2-6m_\pi^2)]+4(t_1-m_\pi^2)(t_2-m_\pi^2) 
\over(s+t_2-2m_\pi^2)^2}\nonumber \\ && +{2s_1(s_1+8m_\pi^2)(t_1-m_\pi^2)(m_\pi^2-t_2)
\over (s+t_2-2m_\pi^2)^3} \bigg\}\Big[J(s_1)- J(2m_\pi^2+s_1-s-t_2)\Big]\Bigg\} 
\,. \end{eqnarray} 
In the process of evaluation one encounters loop integrals over three
pion-propagators with tensors up to third rank ($l_\mu,\,l_\mu l_\nu,\,l_\mu
l_\nu l_\rho$) in the numerator. After the pertinent tensor reduction all the 
occurring scalar loop functions can be expressed as linear combinations of 
$J(\dots)$ and $G(\dots)$ with rational coefficient functions. Note that the 
quality $A_1=A_2$ does no more hold for diagram (V) since the four-momenta $q_1$ 
and $q_2$ of the two neutral pions appear in the loop in a non-symmetrical way. 
The permutational symmetry between both $\pi^0$ is restored by the
contribution from diagram (VI):   
\begin{equation}A_1^{(\rm VI)} = A_2^{(\rm  V)}\Big|(s_1\leftrightarrow 
s_2,\, t_1\leftrightarrow t_2) \,,\qquad  A_2^{(\rm VI)} = A_1^{(\rm  V)}\Big|(s_1
\leftrightarrow s_2,\, t_1\leftrightarrow t_2) \,.  \end{equation}  
 
\begin{figure}
\begin{center}
\includegraphics[scale=1.,clip]{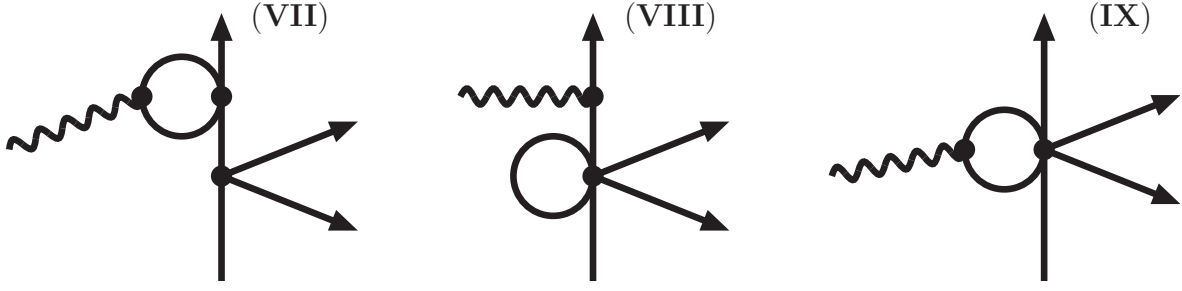}
\end{center}
\vspace{-.5cm}
\caption{One-pion loop diagrams for $\pi^-\gamma  \to \pi^- \pi^0\pi^0$. Their 
combinatoric factor is 1/2.}
\end{figure}

Diagram (VII) in Fig.\,4 generates a (constant) vertex correction to the 
pion-photon coupling and therefore leads to the amplitudes: 
\begin{equation}A_1^{(\rm VII)}=A_2^{(\rm VII)} = {2m_\pi^2\over (4\pi f_\pi)^2}{2m_\pi^2+s
-s_1-s_2 \over 3m_\pi^2-s-t_1-t_2}\bigg(\xi + \ln{m_\pi \over\lambda} \bigg)
\,, \end{equation}
which are proportional to the tree amplitudes written in eq.(5). The chiral 
six-pion vertex appearing in diagrams (VIII) and (IX) represents some challenge 
with respect to tackling the combinatorics involved. Altogether, one finds from 
diagram (VIII) the momentum-dependent amplitudes:  
\begin{equation}A_1^{(\rm VIII)}=A_2^{(\rm VIII)} = {m_\pi^2\over (4\pi f_\pi)^2}
{31m_\pi^2+2(4s-5s_1-5s_2-t_1-t_2)\over 3m_\pi^2-s-t_1-t_2}\bigg(\xi + \ln{m_\pi 
\over\lambda} \bigg)\,, \end{equation}
while diagram (IX) leads to constant amplitudes: 
\begin{equation}A_1^{(\rm IX)}=A_2^{(\rm IX)} =- {2m_\pi^2\over (4\pi f_\pi)^2}
\bigg(\xi + \ln{m_\pi \over\lambda} \bigg)\,. \end{equation}

\begin{figure}
\begin{center}
\includegraphics[scale=1.,clip]{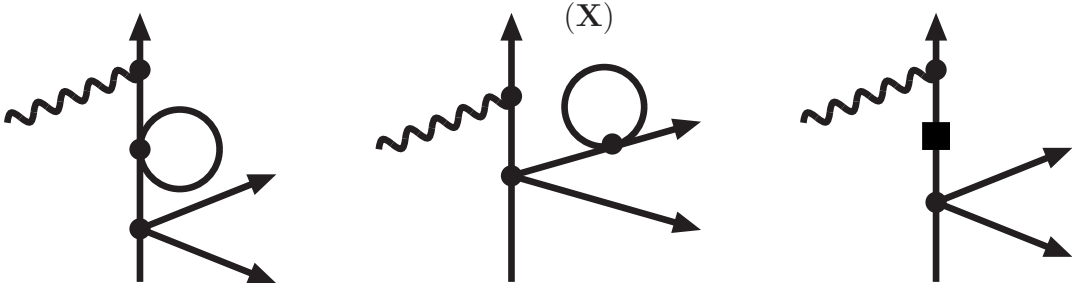}
\end{center}
\vspace{-.5cm}
\caption{Diagrams for $\pi^-\gamma  \to \pi^- \pi^0\pi^0$ involving the pion
wave function renormalization factor. The black square symbolizes the
counterterm proportional to $\ell_4$. Not all possible diagrams of this kind
are shown.}
\end{figure}

Fig.\,5 shows diagrams with self-energy insertions on external or internal 
pion-lines. Their overall effect is to multiply the tree amplitudes 
$A_{1,2}^{(\rm tree)}$ with three times the pion wave function renormalization factor 
$Z_\pi-1$ (see eq.(31) in ref.\cite{pipin}):
\begin{equation}A_1^{(\rm X)}=A_2^{(\rm X)} ={6m_\pi^2\over (4\pi f_\pi)^2}{s_1+s_2 -s- 
2m_\pi^2 \over 3m_\pi^2-s-t_1-t_2}\bigg(3\xi + 16\pi^2 \ell_4^r + \ln{m_\pi\over
\lambda} \bigg)\,. \end{equation}
Note that we have combined the contribution from the pion-loop proportional to
$\xi+\ln(m_\pi/\lambda)$ with that from the counterterm proportional to
$\ell_4=\ell^r_4+ \xi/8\pi^2$.
\begin{figure}
\begin{center}
\includegraphics[scale=1.,clip]{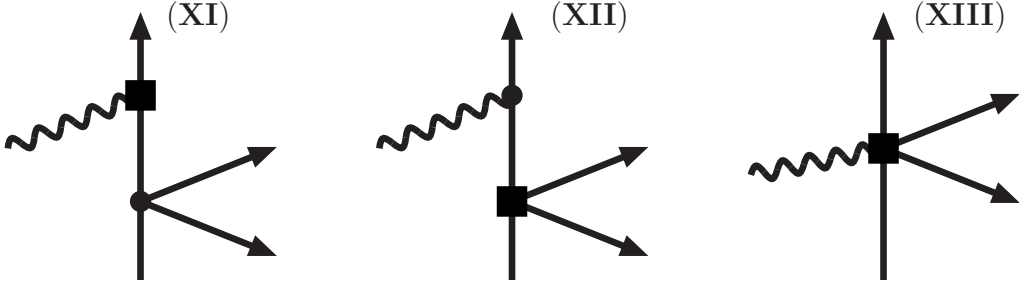}
\end{center}
\vspace{-.5cm}
\caption{Next-to-leading order tree diagrams for $\pi^-\gamma \to \pi^- \pi^0\pi^0$.
 The black square symbolizes the chiral-invariant counterterm proportional to 
$\ell_{1,2,4}$.}
\end{figure}

Fig.\,6 shows the remaining tree diagrams for $\pi^-\gamma \to \pi^- \pi^0\pi^0$
generated by the chiral counterterm Lagrangian ${\cal L}_{\pi\pi}^{(4)}$. We use 
the form of ${\cal L}_{\pi\pi}^{(4)}$ as written in eq.(20) of ref.\cite{pipin}.
Diagram (XI) with a vertex correction at the pion-photon coupling gives rise 
to the amplitudes:   
\begin{equation}A_1^{(\rm XI)}=A_2^{(\rm XI)} = {4m_\pi^2\over (4\pi  f_\pi)^2}{2m_\pi^2+s
-s_1-s_2  \over 3m_\pi^2-s-t_1-t_2}\Big(\xi+8\pi^2\ell_4^r \Big)\,,\end{equation}
which are obviously proportional to the tree amplitudes. Diagram (XII)
involves the four-pion vertex from ${\cal L}_{\pi\pi}^{(4)}$. For our choice of the 
interpolating pion field it has no term proportional to the low-energy constant 
$\ell_3$. With that simplification one finds from diagram (XII):
\begin{eqnarray}A_1^{(\rm XII)}=A_2^{(\rm XII)}&=&{2\over (4\pi  f_\pi)^2}{1\over 3m_\pi^2
-s-t_1-t_2}\bigg\{\bigg({\xi\over 3}+16\pi^2 \ell_1^r\bigg)(s+m_\pi^2-s_1-s_2)
\nonumber \\ &&\times(2s-2m_\pi^2-s_1-s_2+t_1+t_2)+\bigg({\xi\over 3}+8\pi^2
\ell_2^r\bigg)\Big[s_1(s_1+t_1-t_2)\nonumber \\ &&  +s_2(s_2+t_2-t_1)-s
(s_1+s_2+t_1+t_2)+3m_\pi^2(2s-s_1-s_2+t_1+t_2)\nonumber \\ && -2t_1 t_2\Big]+
\Big(\xi+8\pi^2 \ell_4^r\Big)m_\pi^2 (5s+5m_\pi^2-4s_1-4s_2+t_1+t_2)\bigg\}\,. 
\end{eqnarray} 
Finally, there is diagram (XIII) involving the (next-to-leading order) 
$\gamma 4\pi$ contact vertex from  ${\cal L}_{\pi\pi}^{(4)}$. Its polynomial 
contribution to the amplitudes: 
\begin{eqnarray}A_1^{(\rm XIII)}&=&{2\over (4\pi f_\pi)^2}\bigg\{\bigg({\xi\over 3}
+16\pi^2 \ell_1^r\bigg)(s+m_\pi^2-s_1-s_2)  \nonumber \\ &&  +\bigg({\xi\over  3}
+8\pi^2\ell_2^r\bigg)(2m_\pi^2-s+s_1-s_2-t_2) +\Big(\xi+8\pi^2 \ell_4^r\Big)m_\pi^2  
\bigg\}\,, \end{eqnarray} 
\begin{eqnarray}A_2^{(\rm XIII)}&=&{2\over (4\pi f_\pi)^2}\bigg\{\bigg({\xi\over 3}
+16\pi^2 \ell_1^r\bigg)(s+m_\pi^2-s_1-s_2)  \nonumber \\ &&  +\bigg({\xi\over  3}
+8\pi^2\ell_2^r\bigg)(2m_\pi^2-s-s_1+s_2-t_1) +\Big(\xi+8\pi^2 \ell_4^r\Big)m_\pi^2  
\bigg\}\,, \end{eqnarray} 
is exceptional in the sense that $A_1$ and $A_2$ are not equal. Up to that point
the compilation of the production amplitudes  $A_{1,2}$ for the process 
$\pi^-\gamma\to\pi^-\pi^0\pi^0$ as they arise from chiral loops and counterterms
is completed. There is still one issue which has to addressed, namely the
renormalization of the squared pion decay constant $f_\pi^2$ and squared pion 
mass $m_\pi^2$ by chiral loops and counterterms (see herefore e.g. eqs.(29,30)
in  ref.\cite{pipin}). The chiral Lagrangian ${\cal L}_{\pi\pi}$ operates with 
their leading-order values whereas the tree amplitudes in eqs.(3,5) have been 
written already in terms their physical values. In order to correct for this 
difference of order $(m_\pi/2\pi f_\pi)^2$ one has to add an extra
renormalization contribution of the form:     
\begin{eqnarray}A_1^{(\rm ren)}=A_2^{(\rm ren)} &=&{m_\pi^2\over (4\pi f_\pi)^2}{1 \over 
3m_\pi^2-s-t_1-t_2}\bigg\{4(2m_\pi^2+s-s_1-s_2)\nonumber \\ &&\times \bigg(8\pi^2
\ell_4^r -\ln{m_\pi \over \lambda} \bigg)+ m_\pi^2 \bigg(32\pi^2\ell_3^r +\ln{m_\pi 
\over \lambda} \bigg)\bigg\}\,.\end{eqnarray}
A first crucial check of our calculation is provided by the fact that the
ultraviolet divergence $\xi$ drops out in the total sums for $A_1$ and $A_2$.
Next, one can further simplify the expressions for the amplitudes by 
introducing via the relation: 
\begin{equation}  \ell_j^r = {\gamma_j \over 32\pi^2} \bigg( \bar \ell_j +
  2\ln{m_\pi \over \lambda}\bigg)\,, \qquad \gamma_1 = {1\over 3}\,,\quad 
\gamma_2 = {2\over 3}\,,\quad  \gamma_3 = -{1\over 2}\,,\quad \gamma_4 = 2\,,
\end{equation}
the ''barred'' low-energy constants $\bar \ell_j$ which subsume the chiral 
logarithm $\ln(m_\pi/\lambda)$. After summing up the expressions in eqs.(18-26) 
and adding the terms proportional to the chiral logarithm $\ln(m_\pi/\lambda)$ 
in the loop amplitudes eqs.(6,10,11,12,15,16,17) one gets the following 
modified and complete counterterm contributions: 
\begin{eqnarray}A_1^{(\rm ct)}&=&{1\over (4\pi f_\pi)^2}{1 \over 3m_\pi^2-s-t_1-t_2}
\bigg\{{\bar \ell_1 \over 3} (s_1+s_2-s-m_\pi^2)^2 +{\bar \ell_2 \over 3} \Big[
s^2+s_1^2+s_2^2 \nonumber \\ &&+t_2^2-2s s_1+(s-2s_1+2s_2-t_1)t_2+m_\pi^2(s-6s_2+t_1
-2t_2+6m_\pi^2) \Big] \nonumber \\ &&- {\bar \ell_3 \over 2}m_\pi^4 +2\bar \ell_4
 m_\pi^2(s+2m_\pi^2-s_1-s_2) \bigg\} \,,\end{eqnarray}
\begin{eqnarray}A_2^{(\rm ct)}&=&{1\over (4\pi f_\pi)^2}{1 \over 3m_\pi^2-s-t_1-t_2}
\bigg\{{\bar \ell_1 \over 3} (s_1+s_2-s-m_\pi^2)^2 +{\bar \ell_2 \over 3} \Big[
s^2+s_1^2+s_2^2 \nonumber \\ &&+t_1^2-2s s_2+(s+2s_1-2s_2-t_2)t_1+m_\pi^2(s-6s_1+t_2
-2t_1+6m_\pi^2) \Big] \nonumber \\ &&- {\bar \ell_3 \over 2}m_\pi^4 +2\bar \ell_4
 m_\pi^2(s+2m_\pi^2-s_1-s_2) \bigg\} \,.\end{eqnarray}
The remaining finite parts of the loop amplitudes are then given by
eqs.(6,10,11,12,15,16,17) with the $\xi+ \ln(m_\pi/\lambda)$ terms deleted
altogether. 

Let us also mention that the other three terms in the chiral Lagrangian ${\cal L
}_{\pi\pi}^{(4)}$ proportional to the low-energy constants $\ell_{5,6,7}$ \cite{buergi} 
do not contribute to the $\pi^- \gamma \to 3\pi$ processes considered in this work.
 The $\ell_7$ term breaks isospin symmetry and is ignored for this reason. The 
$l_5$ term requires (for electromagnetic process) at least two external photons.
The $\ell_6$ term gives rise to a correction to the pion-photon coupling, which 
however vanishes for real photons ($k^2=0=\epsilon \cdot k$). Let us remind 
that $\ell_6$ is related to the mean square charge radius of the pion via: 
$\langle r^2_\pi\rangle = (\bar \ell_6-1)/(4\pi f_\pi)^2 +{\cal O}(m_\pi^2)$ 
\cite{gasleut}. The associated $\gamma 4\pi$ vertex vanishes also since the
commutator in tr$(\tau_3[\partial_\mu U,\partial_\nu U^\dagger])$ terminates at the
quadratic order in the pion-field (for our choice of interpolating pion-field).

\subsection{Results for $\pi^- \gamma\to\pi^-\pi^0\pi^0$}
In this section we present and discuss the results for the neutral pion-pair
production process  $\pi^-\gamma\to\pi^-\pi^0\pi^0$ at next-to-leading order
in chiral perturbation theory. We use for the low-energy constants $\bar\ell_j$
the values: $\bar \ell_1 = -0.4\pm 0.6$, $\bar \ell_2 = 4.3\pm 0.1$, $\bar\ell_3
=2.9\pm 2.4$, $\bar \ell_4 =4.4\pm 0.2$, as determined (with improved empirical
input) in ref.\cite{cola}. Applying the flux and symmetry factors the total 
cross section $\sigma_{\rm tot}(s)$ is obtained by integrating the squared
(transversal) T-matrix over the three-pion phase space:  
\begin{equation}\sigma_{\rm tot}(s)={\alpha \over 32\pi^3 f_\pi^4 (s-m_\pi^2)} 
\int\limits_{z^2<1}\!\!\!\!\!\int\!d\omega_1 d\omega_2 \int_{-1}^1\!dx\int_0^\pi \! 
d\phi \, \Big|\hat k \times (\vec q_1 A_1+\vec q_2 A_2)\Big|^2 \,.\end{equation} 
Here, $\omega_{1,2}$ are the center-of-mass energies of the out-going neutral
pions and $q_{1,2} = \sqrt{\omega_{1,2}^2 -m_\pi^2}$. In terms of the directional 
cosines  $x,y,z$ the squared cross products in eq.(30) take the form: 
\begin{equation} (\hat k \times \vec q_1)^2 = q_1^2(1-x^2)\,, \quad  (\hat k
\times \vec q_2)^2 = q_2^2(1-y^2)\,, \quad   (\hat k \times \vec q_1)\cdot 
(\hat k \times \vec q_2)= q_1 q_2(z-x y)\,, \end{equation}
together with the relations: 
\begin{equation} q_1q_2 \,z =\omega_1 \omega_2-\sqrt{s}(\omega_1 +\omega_2) 
+{s+m_\pi^2 \over 2}  \,, \quad\quad y = xz +\sqrt{(1-x^2)(1-z^2)} \cos\phi \,.
\end{equation}
The Mandelstam variables $s_1,\,s_2,\,t_1,\,t_2$ follow as:
\begin{eqnarray} && s_1 = s+m_\pi^2-2 \sqrt{s}\, \omega_2\,, \quad  s_2 =  
s+m_\pi^2-2 \sqrt{s} \,\omega_1\,,\nonumber \\ && t_1 = m_\pi^2+{m_\pi^2-s\over 
\sqrt{s}} (\omega_1-q_1 x) \,,\quad t_2 = m_\pi^2+{m_\pi^2-s\over \sqrt{s}} 
(\omega_2-q_2 y)\,. \end{eqnarray}
The more definite expression of the tree-level cross section given in eq.(2) is 
very helpful for checking the numerical accuracy of the four-dimensional
integration involved in eq.(30).

\begin{figure}
\begin{center}
\includegraphics[scale=.47,clip]{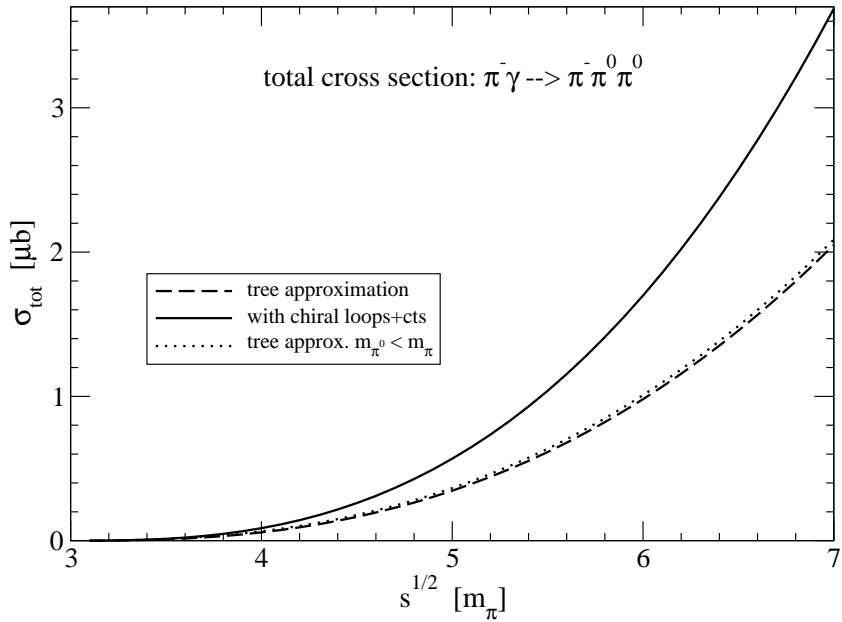}
\end{center}
\vspace{-.7cm}
\caption{Total cross section for the reaction $\pi^- \gamma \to \pi^-\pi^0\pi^0$
as a function of the center-of-mass energy $\sqrt{s}$.}
\end{figure}

Fig.\,7 show the total cross section $\sigma_{\rm tot}(s)$ for the reaction $\pi^-
\gamma\to\pi^-\pi^0\pi^0$ in the low-energy region from threshold $\sqrt{s}=
3m_\pi$ up to  $\sqrt{s}=7m_\pi$. The dashed line corresponds to the tree
approximation and the full line includes in addition the next-to-leading
order corrections from chiral loops and counterterms. The dotted line follows
if the charged and neutral pion mass are distinguished ($134.977\,{\rm MeV}
=m_{\pi^0}< m_\pi = 139.570\,{\rm MeV}$) in the tree-level amplitude and the 
three-pion phase space integral (see eq.(2)). As it could be expected such 
isospin-breaking effects are very small. By inspection of Fig.\,7 one observes 
that the total cross section for $\pi^- \gamma\to\pi^-\pi^0\pi^0$ gets enhanced 
sizeably (by a factor of $1.5-1.8$) after inclusion of the next-to-leading order 
chiral corrections. Although the dynamics of the whole process is much richer
this feature can be understood (in an approximate way) from the $\pi^+\pi^-\to
\pi^0\pi^0$ final state interaction. The  $\pi^+\pi^- \to \pi^0\pi^0$ interaction
strength at threshold is determined by the difference of the isospin-zero and 
isospin-two S-wave $\pi\pi$-scattering lengths. When considering the 
corresponding  one-loop expression \cite{gasleut}:  
\begin{equation} {1\over 3}(a_0^0-a_0^2) = {3m_\pi \over 32\pi  f_\pi^2}\bigg[ 1
+{m_\pi^2 \over 36 \pi^2 f_\pi^2} \bigg( \bar \ell_1+2 \bar \ell_2-{3  \bar\ell_3
\over 8}+{9 \bar \ell_4\over 2} +{33\over 8} \bigg) \bigg] \,, \end{equation}
one sees that the correction to $1$ inside the square bracket amounts to about 
$0.20$ (inserting the central values of $\bar \ell_j$). Indeed, the square $1.2^2
=1.44$ is close to the enhancement factor $1.5$ of the total cross section at 
$\sqrt{s} =4m_\pi$. 

It is also important to investigate the uncertainties which are induced by the
present errorbars $\delta\bar \ell_j$ of the low-energy constants $\bar\ell_j$.
Taking the total cross section at $\sqrt{s} =6m_\pi$ as a measure one finds
relative uncertainties of: $\pm 5.1\%$ from $\delta\bar \ell_1$,  $\pm 1.4\%$ 
from $\delta\bar \ell_2$, $\pm 0.2\%$ from $\delta\bar \ell_3$, and $\pm
0.9\%$ from $\delta\bar \ell_4$. It is comforting that the badly known
low-energy constant $\bar \ell_3 = 2.9\pm 2.4$ has very little influence on
the observables considered here. The largest uncertainty is actually connected 
with $\bar \ell_1$ and adding them all in quadrature one gets a total relative 
uncertainty of $\pm 5.4\%$. This amounts to a fairly accurate prediction.      

In addition to the total cross section there are more exclusive observables 
such as the two-pion mass spectra. The $\pi^0\pi^0$ invariant mass $\mu$ is 
defined by $\mu^2 =(q_1+q_2)^2 = s-s_1-s_2+3m_\pi^2$ and it varies over the 
kinematically allowed  range $2m_\pi <\mu <\sqrt{s}-m_\pi$. In tree approximation
the $\pi^0\pi^0$ mass spectrum can be given in closed analytical form:
\begin{equation} {d\sigma \over d\mu } = {\alpha \sqrt{\mu^2-4m_\pi^2} \over
16 \pi^2 f_\pi^4(s-m_\pi^2)^3} (\mu^2-m_\pi^2)^2 \bigg\{(s+m_\pi^2-\mu^2)\ln{s+m_\pi^2
-\mu^2 +\sqrt{W} \over 2m_\pi \sqrt{s}} - \sqrt{W}\bigg\} \,,\end{equation} 
with the abbreviation $W= [s-(\mu+m_\pi)^2][s-(\mu-m_\pi)^2]$. In order to
obtain $d\sigma/d\mu$ in general one introduces new energy variables $\omega_+
= \omega_1+ \omega_2$ and $\omega_-=(\omega_1- \omega_2)/2$ such that $\mu^2=
2 \sqrt{s}\, \omega_++m_\pi^2-s$. The $d\omega_+$ integration in eq.(30) is 
omitted and a normalization factor $\mu /\sqrt{s}$ is  applied. The lower and
upper limit for the $d\omega_-$ integration are $\mp\sqrt{(\mu^2-4m_\pi^2)W}/
(4\mu \sqrt{s})$. Fig.\,8 shows the calculated $\pi^0\pi^0$ mass spectrum
$m_\pi\,d\sigma/d\mu$. We have multiplied it by the constant factor $m_\pi$ in 
order to keep the units ($\mu$b) of a cross section. The four pairs of (full 
and dashed) curves correspond to sections at center-of-mass-energies $\sqrt{s}
=(4,5,6,7)m_\pi$ in ascending order. In essence Fig.8 reproduces the enhancement
of the total cross section by the next-to-leading order chiral corrections. No 
further specific dynamical details to which the $\pi^0\pi^0$ mass spectrum could
be selectively sensitive are visible.
      
\begin{figure}
\begin{center}
\includegraphics[scale=.47,clip]{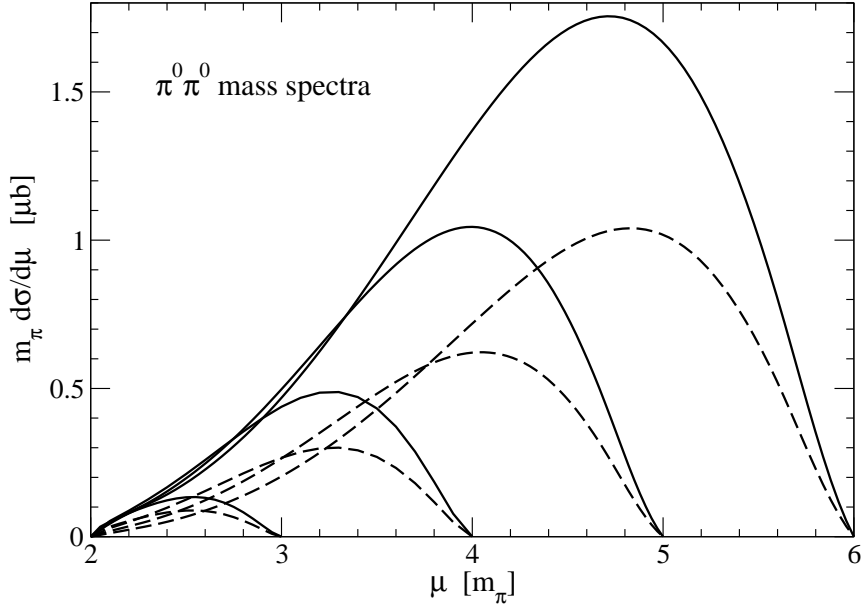}
\end{center}
\vspace{-.7cm}
\caption{$\pi^0\pi^0$ mass spectrum for the reaction $\pi^- \gamma \to\pi^-\pi^0
\pi^0$ as a function of the $\pi^0\pi^0$ invariant mass $\mu$. The curves 
correspond to center-of-mass energies $\sqrt{s}=(4,5,6,7)m_\pi$ in ascending order.}
\end{figure}

The alternative combination is to couple one of neutral pions with the 
out-going $\pi^-$. The $\pi^0\pi^-$ invariant mass is $\sqrt{s_1}$ (or
equivalently $\sqrt{s_2}$\,). Making use of the relation $s_1=s+m_\pi^2-2\sqrt{s}
\,\omega_2$ the $\pi^0\pi^-$ mass spectrum $d\sigma/d\sqrt{s_1}$ is obtained by 
omitting the $d\omega_2$ integration in eq.(30) and applying an additional 
factor $\sqrt{s_1/s}$. Fig.\,9 shows the calculated $\pi^0\pi^-$ mass spectrum 
$m_\pi\,d\sigma/d\sqrt{s_1}$. Again, it is only the enhancement of the total 
cross section which can be inferred from the comparison of the full and dashed 
curves in Fig.\,9. The shape of the $\pi^0\pi^-$ mass  spectrum (i.e. its 
dependence on $\sqrt{s_1}$\,) does not distinguish the  tree-approximation
from the full calculation in a noticeable way. 

\begin{figure}
\begin{center}
\includegraphics[scale=.47,clip]{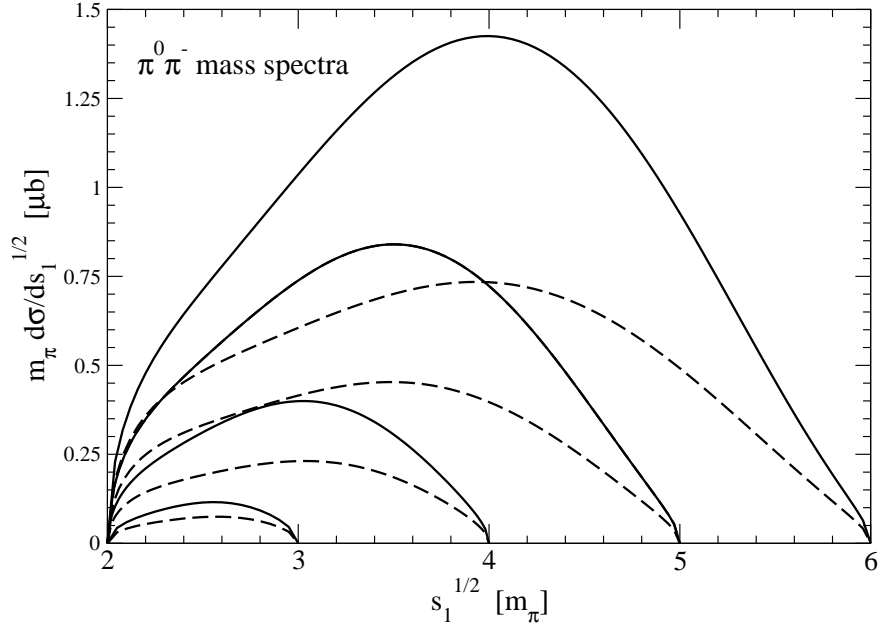}
\end{center}
\vspace{-.7cm}
\caption{$\pi^0\pi^-$ mass spectrum for the reaction $\pi^- \gamma \to\pi^-\pi^0
\pi^0$ as a function of the $\pi^0\pi^-$ invariant mass $\sqrt{s_1}$. The curves 
correspond to center-of-mass energies $\sqrt{s}=(4,5,6,7)m_\pi$ in ascending order.}
\end{figure}

\section{ Charged pion-pair production}
In this section we perform the same calculation and analysis for the charged 
pion-pair production process: $\pi^-(p_1)+\gamma(k,\epsilon\,) \to\pi^+(p_2)+
\pi^-(q_1)+\pi^-(q_2)$. By assigning the four-momentum $p_2$ to the out-going 
positively charged pion $\pi^+(p_2)$  we can exploit the complete equivalence 
to the $\pi^-\gamma\to \pi^- \pi^0 \pi^0$ reaction concerning its kinematical 
description. In Coulomb-gauge ($\epsilon\cdot p_1= \epsilon\cdot k=0$), eq.(3) 
constitutes the general form of the T-matrix for $\pi^-\gamma\to \pi^+ \pi^- 
\pi^-$ and the corresponding Mandelstam variables are defined as in eq.(4). The
interchange of the two identical $\pi^-$ in the final state is now described 
by $(s_1\leftrightarrow s_2,\, t_1\leftrightarrow t_2)$. The three non-vanishing 
tree diagrams for $\pi^- \gamma\to \pi^+ \pi^- \pi^-$ are shown in Fig.\,1 and 
their evaluation leads to the following tree amplitudes:
\begin{equation}A_1^{(\rm tree)} = {s+m_\pi^2-s_1-s_2\over 3m_\pi^2-s-t_1-t_2} +{s-s_1
-s_2+t_2 \over t_1-m_\pi^2}-1 \,, \end{equation}
\begin{equation}A_2^{(\rm tree)} = {s+m_\pi^2-s_1-s_2\over 3m_\pi^2-s-t_1-t_2} +{s-s_1
-s_2+t_1 \over t_2-m_\pi^2}-1 \,. \end{equation}
One can see from the denominators of $A_1^{(\rm tree)}$ and $A_2^{(\rm tree)}$ that two
diagrams contribute to each amplitude. 
\subsection{Amplitudes from chiral loops and counterterms}
Beyond leading order the dynamical content of charged pion-pair production $\pi^-
\gamma\to \pi^+ \pi^- \pi^-$ is considerably more extensive than that of neutral 
pion-pair production $\pi^- \gamma\to\pi^- \pi^0 \pi^0 $ because the photon can now 
couple to all three out-going (charged) pions. Many more diagrams with chiral 
pion-loops and counterterms do contribute. We have evaluated them individually
and checked the exact cancellation of ultraviolet divergences $\xi$ in the
total sums for the amplitudes $A_1$ and $A_2$. Without loss of information we can 
restrict the presentation of the analytical results to the finite parts of the 
pion-loop diagrams and to the complete counterterm contribution (reexpressed in 
terms of the low-energy constants $\bar \ell_j$ which subsume the chiral 
logarithm $\ln(m_\pi/\lambda)$ generated by the pion-loops).

\begin{figure}
\begin{center}
\includegraphics[scale=1.,clip]{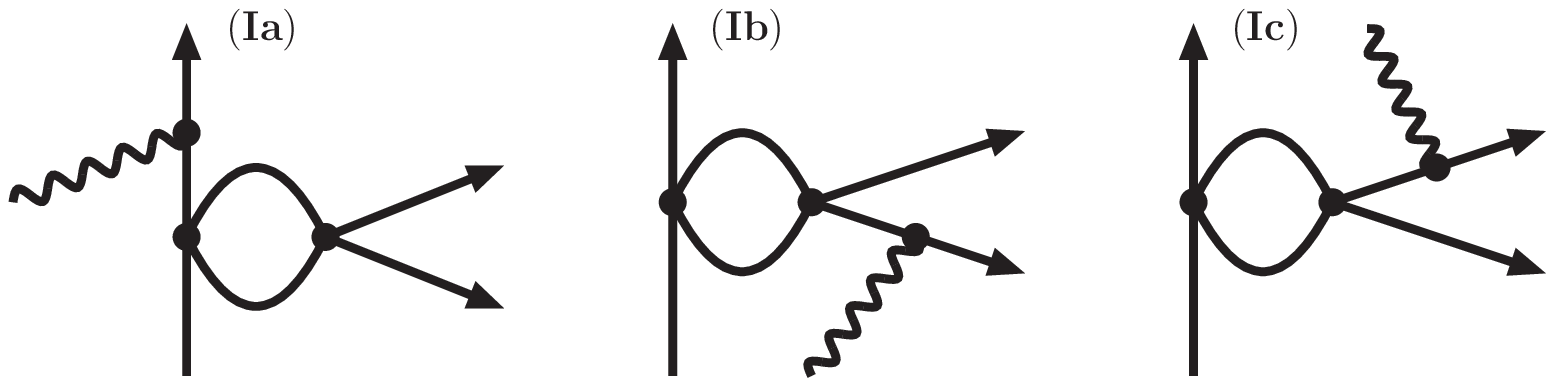}
\end{center}
\vspace{-.5cm}
\caption{One-pion loop diagrams for $\pi^-\gamma\to \pi^+  \pi^- \pi^-$ with
three possible couplings of the external photon.}
\end{figure}

The three possible couplings of the external photon for loop diagram (I) (see 
Fig.\,2) are distinguished by labelling them (Ia), (Ib), (Ic) and are shown in 
Fig.\,10. Omitting the terms proportional to $\xi+\ln(m_\pi/\lambda)$ the
corresponding finite parts read:
\begin{eqnarray} A_1^{(\rm Ia)}=A_2^{(\rm Ia)}&=&{1\over(4\pi f_\pi)^2}{s+m_\pi^2-s_1-s_2 
\over 3m_\pi^2-s-t_1-t_2}(2s-2m_\pi^2-s_1-s_2+t_1+t_2)\nonumber \\ && \times \bigg[ 
J(3m_\pi^2+s-s_1-s_2)-{1\over 2}\bigg]  \,,\ \end{eqnarray} 
\begin{eqnarray} A_1^{(\rm Ib)} &=&{1\over(4\pi f_\pi)^2}{s-s_1-s_2+t_2   \over t_1- 
m_\pi^2} (s-m_\pi^2-s_1-s_2+t_1+t_2)\nonumber \\ && \times \bigg[ J(m_\pi^2+s-s_1
-s_2+t_1+t_2)-{1\over 2}\bigg]  \,,\ \end{eqnarray} 
\begin{equation}A_2^{(\rm Ic)} = A_1^{(\rm Ib)}\Big|(s_1\leftrightarrow s_2, \, t_1
\leftrightarrow t_2) \,.   \end{equation}
\begin{figure}
\begin{center}
\includegraphics[scale=1.,clip]{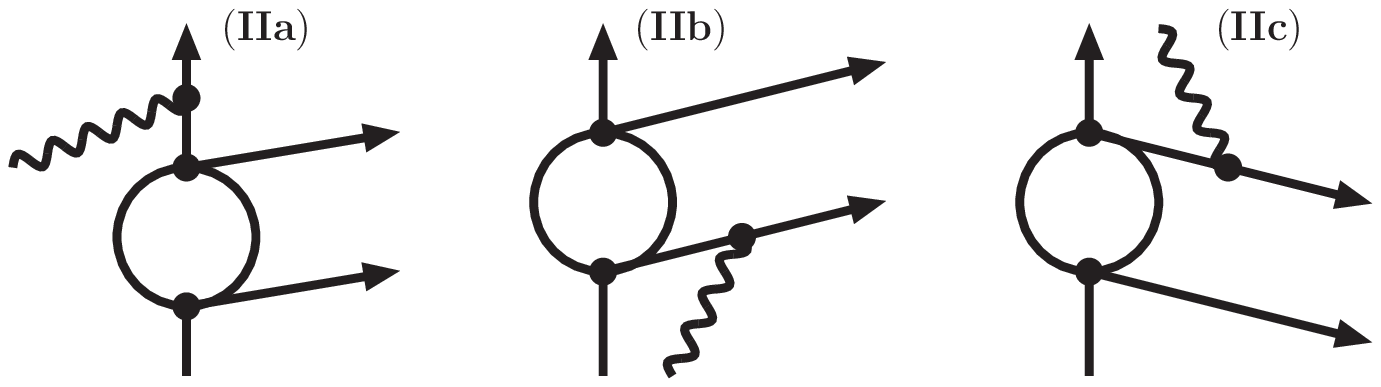}
\end{center}
\vspace{-.5cm}
\caption{One-pion loop diagrams for $\pi^-\gamma\to \pi^+  \pi^- \pi^-$ with
three possible couplings of the external photon.}
\end{figure}

Note that we do not list vanishing contributions to $A_1$ or $A_2$ from a 
diagram under consideration. Fig.\,11 shows the one-pion loop diagrams (IIa), 
(IIb), (IIc) obtained from the rescattering diagram (II) (see Fig.\,2) by
attaching the external photon in the three possible ways. One finds the 
following finite parts:
\begin{eqnarray} A_1^{(\rm IIa)}=A_2^{(\rm IIa)}&=&{1\over 3(4\pi  f_\pi)^2}{1\over 
3m_\pi^2-s-t_1-t_2}\bigg\{{m_\pi^2\over 3}(21s+7s_1-20s_2+28t_1+t_2)\nonumber \\ &&
+{1\over 6}(s-s_2+t_1)(5s_1+16s_2-23s-18t_1-7t_2)-{61m_\pi^4\over6}\nonumber \\ && 
+\Big[m_\pi^2(19m_\pi^2-17s-2s_1 +15s_2-19t_1 -2t_2)+(s-s_2+t_1)\nonumber \\ &&
\times (7s-s_1-5s_2+6t_1+2t_2)\Big]J(2m_\pi^2-s+s_2-t_1)\bigg\}\,, \end{eqnarray} 
\begin{eqnarray} A_1^{(\rm IIb)}&=&{1\over 3(4\pi  f_\pi)^2}{1\over  t_1-m_\pi^2}
\bigg\{{m_\pi^4\over 2}+{2m_\pi^2\over 3}(6s_1+8s_2-6s-3t_1-6t_2)\nonumber \\ &&
+{s_2\over 6}(5s-5s_1-16s_2-2t_1+5t_2)+\Big[s_2(5s_2+s_1+t_1-s-t_2)\nonumber \\ &&
+m_\pi^2(4s+m_\pi^2-4s_1 -9s_2+2t_1+4t_2)\Big] J(s_2) \bigg\}\,, \end{eqnarray} 
\begin{eqnarray} A_2^{(\rm IIc)}&=&{1\over 3(4\pi  f_\pi)^2}{1\over  t_2-m_\pi^2}
\bigg\{{m_\pi^2 \over 3}(11s+7s_1-11s_2+11t_1-8t_2)\nonumber\\ &&-{25m_\pi^4\over 6}
+{1\over 6}(s-s_2+t_1)(5s_1+16s_2+2t_2-16s-16t_1)\nonumber \\ &&+\Big[m_\pi^2(9s_2
+4t_2-9s-2s_1-9t_1+7m_\pi^2) +(s-s_2+t_1)\nonumber \\ &&\times
(5s-s_1-5s_2+5t_1-t_2)\Big] J(2m_\pi^2-s+s_2-t_1)\bigg\}\,. \end{eqnarray} 
The additional contributions from the loop diagrams (IIIa), (IIIb), (IIIc) with
crossed out-going $\pi^-$ lines (see Fig.\,2) follow immediately via the 
substitution $q_1 \leftrightarrow q_2$ as: 
\begin{equation}A_1^{(\rm IIIa)} = A_2^{(\rm IIIa)} = A_1^{(\rm  IIa)}\Big|(s_1
\leftrightarrow s_2, \, t_1\leftrightarrow t_2) \,,   \end{equation}
\begin{equation}A_1^{(\rm IIIb)} = A_2^{(\rm IIc)}\Big|(s_1\leftrightarrow  s_2,\, t_1
\leftrightarrow t_2) \,, \qquad A_2^{(\rm IIIc)} = A_1^{(\rm IIb)}\Big|(s_1
\leftrightarrow s_2,\,  t_1\leftrightarrow t_2) \,. \end{equation}
Next, we come to the irreducible one-pion loop diagrams (VI), (V), (VI) with
internal photon coupling shown in Fig.\,3 and interpreted now as diagrams for 
$\pi^-\gamma \to \pi^+  \pi^- \pi^-$. One finds the following finite parts:
\begin{eqnarray} A_1^{(\rm IV)}=A_2^{(\rm  IV)}&=&{2(s-s_1-s_2)+t_1+t_2\over(4\pi f_\pi
)^2}\bigg\{{1\over 2}+{1\over 2m_\pi^2-t_1-t_2 }\nonumber \\ && \times \Big[
(s-m_\pi^2-s_1-s_2+t_1+t_2) \,J(m_\pi^2+s-s_1-s_2+t_1+t_2)\nonumber \\ && 
+ (s_1+s_2-s-m_\pi^2) \,J(3m_\pi^2+s-s_1-s_2) \Big]\bigg\}\,, \end{eqnarray} 
\begin{eqnarray} A_1^{(\rm V)}&=&{1\over 3(4\pi f_\pi)^2}\bigg\{10m_\pi^2-3s-3t_2
+{13s_1\over 6}+6m_\pi^2\bigg(1-{s_1\over s+t_2-2m_\pi^2}\bigg)\nonumber \\ &&
\times \Big[G(2m_\pi^2-s+s_1-t_2)-G(s_1)\Big] +(3s-2s_1+3t_2-10m_\pi^2)\, J(s_1)
\nonumber \\ && +{3(s-s_1+t_2-2m_\pi^2)^2\over s+t_2-2m_\pi^2}\Big[
J(2m_\pi^2-s+s_1-t_2) -J(s_1)\Big] \bigg\}\,, \end{eqnarray}  
\begin{eqnarray} A_2^{(\rm V)}&=&{1\over 3(4\pi f_\pi)^2}\bigg\{{1\over 6}(19m_\pi^2
+13s_1+9t_1-2t_2-2s)+{3s_1(2m_\pi^2-t_1-t_2)\over 2(s+t_2-2m_\pi^2)}\nonumber \\ &&
+6m_\pi^2\bigg[{m_\pi^2-t_1 \over  s+t_2-2m_\pi^2}+{s_1(t_1-s)\over(s+t_2-2m_\pi^2)^2}
\bigg]\Big[G(2m_\pi^2-s+s_1-t_2)-G(s_1)\Big] \nonumber \\ && +\bigg[s_1+2m_\pi^2+
{3s_1^2(s-t_1) \over (s+t_2-2m_\pi^2)^2}+{3s_1(t_1-s) \over  s+t_2-2m_\pi^2}\bigg]
\Big[J(2m_\pi^2-s+s_1-t_2)\nonumber \\ && -J(s_1)\Big] +(s-2s_1+t_2-6m_\pi^2)\,
J(2m_\pi^2-s+s_1-t_2) \bigg\}\,, \end{eqnarray} 
\begin{equation}A_1^{(\rm VI)} = A_2^{(\rm V)}\Big|(s_1\leftrightarrow  s_2,\, t_1
\leftrightarrow t_2) \,, \qquad A_2^{(\rm VI)} = A_1^{(\rm V)}\Big|(s_1
\leftrightarrow s_2,\,  t_1\leftrightarrow t_2) \,, \end{equation}
where the contributions from diagram (VI) are obtained from those of diagram 
(V) by applying the crossing transformation $q_1 \leftrightarrow q_2$. The set of
next-to-leading order corrections to $\pi^-\gamma \to \pi^+  \pi^- \pi^-$ is
completed by the total counterterms contribution which reads: 
\begin{eqnarray} A_1^{(\rm ct)}&=&{1\over(4\pi f_\pi)^2}\Bigg\{{\bar\ell_1\over 3}
\bigg[s+3m_\pi^2-2s_2+t_1-t_2-{(s-s_1+t_2)^2+(s_2-2m_\pi^2)^2 \over  t_1-m_\pi^2}
\nonumber \\ && -{(s-s_1+t_2)^2+(s_2+t_2-3m_\pi^2)^2 \over 3m_\pi^2-s-t_1-t_2} 
\bigg] +{\bar\ell_2\over 3} \bigg[5s+5m_\pi^2-4s_1-6s_2+3t_1-t_2\nonumber \\ && 
+{4s_2(s+m_\pi^2-s_1+t_2)-3s_2^2-4m_\pi^4 -3(s-s_1+t_2)^2 \over t_1-m_\pi^2} 
+{1\over 3m_\pi^2-s-t_1-t_2}\nonumber \\ &&\times \Big[2(s-s_1)(2s_2+2t_2-5m_\pi^2)
-3(s-s_1+t_2-m_\pi^2)^2+t_2^2-8m_\pi^4 \nonumber \\ &&+s_2(10m_\pi^2-3s_2-2t_2)\Big]
\bigg] +\bar\ell_3 \bigg[{m_\pi^4\over t_1-m_\pi^2}+{m_\pi^4 \over  3m_\pi^2-s-t_1
-t_2}\bigg] \nonumber \\ && +2m_\pi^2 \, \bar\ell_4 \bigg[{s+m_\pi^2-s_1-s_2 
\over  3m_\pi^2-s -t_1-t_2}+{s-s_1 -s_2+t_2 \over t_1-m_\pi^2}-1\bigg]\Bigg\}\,, 
\end{eqnarray} 
\begin{equation}A_2^{(\rm ct)} = A_1^{(\rm ct)}\Big|(s_1\leftrightarrow  s_2,\, t_1
\leftrightarrow t_2) \,.  \end{equation}
The relation $A_2= A_1|(s_1\leftrightarrow  s_2,\, t_1\leftrightarrow t_2)$
holds also for the total sum of the loop amplitudes and it applies to both
reactions  $\pi^-\gamma \to \pi^+  \pi^- \pi^-$ and  $\pi^-\gamma \to \pi^- \pi^0
\pi^0$ in the same way.  
\subsection{Results for $\pi^-\gamma\to\pi^+\pi^-\pi^-$}
We are now in the position to present numerical results for the charged
pion-pair production process $\pi^-\gamma \to \pi^+  \pi^- \pi^-$ at
next-to-leading order in chiral perturbation theory. The formula for calculating
the total cross section $\sigma_{\rm tot}(s)$ is given in unchanged form by 
eq.(30). We use consistently the same values: $\bar \ell_1 = -0.4\pm 0.6$,
$\bar \ell_2 = 4.3\pm 0.1$, $\bar\ell_3=2.9\pm 2.4$, $\bar \ell_4 = 4.4\pm 0.2$,
for the low-energy constants $\bar \ell_j$ as in subsection 2.2. 

\begin{figure}
\begin{center}
\includegraphics[scale=.47,clip]{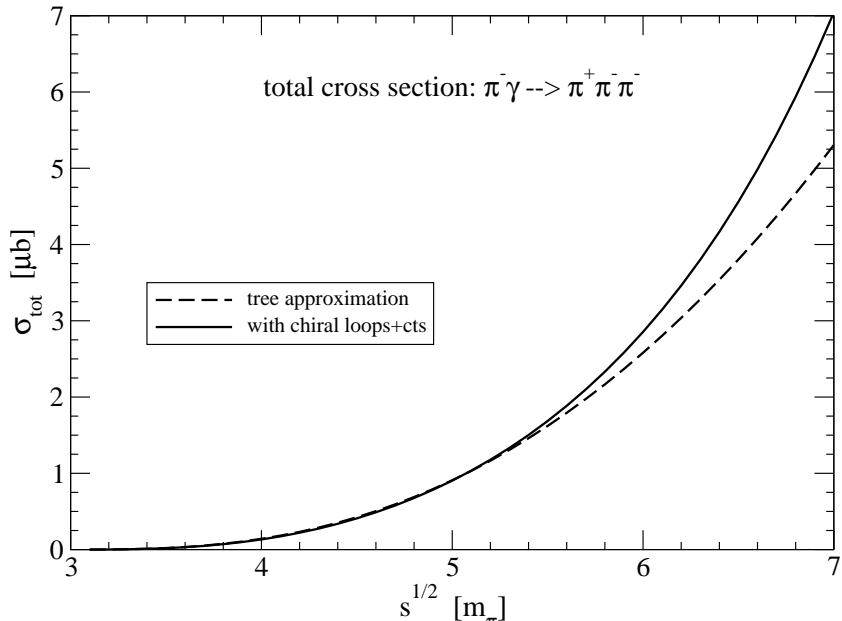}
\end{center}
\vspace{-1.cm}
\caption{Total cross section for the reaction $\pi^- \gamma \to \pi^+\pi^-\pi^-$
as a function of the center-of-mass energy $\sqrt{s}$.}
\end{figure}

Fig.\,12 shows the total cross section for $\sigma_{\rm tot}(s)$ for the reaction 
$\pi^-\gamma\to\pi^+\pi^-\pi^-$ in the low-energy region from threshold $\sqrt{s}=
3m_\pi$ up to  $\sqrt{s}=7m_\pi$. The dashed line corresponds to the tree
approximation and the full line includes in addition the next-to-leading
order corrections from chiral loops and counterterms. By inspection of Fig.\,12 
one observes that the total cross section for $\pi^- \gamma\to\pi^+\pi^-\pi^-$ 
remains almost unchanged in the region $\sqrt{s}<6m_\pi$ after inclusion of 
the next-to-leading order chiral corrections. This striking result is in 
marked contrast to the behavior of the total cross section for neutral 
pion-pair production $\pi^- \gamma\to\pi^-\pi^0\pi^0$ (see Fig.\,7) Although the
dynamics of the whole process is much richer this feature can be understood 
(in a suggestive way) from the $\pi^-\pi^-\to \pi^-\pi^-$ final state interaction. 
By considering the one-loop expression for the isospin-two S-wave $\pi\pi$ 
scattering length \cite{gasleut}: 
\begin{equation} a_0^2 = -{m_\pi \over 16\pi  f_\pi^2}\bigg[ 1-{m_\pi^2 \over 12 
\pi^2 f_\pi^2} \bigg(\bar \ell_1+2 \bar \ell_2-{3\bar \ell_3\over 8}-{3\bar \ell_4
\over 2} +{3\over 8} \bigg) \bigg] \,, \end{equation}
one sees that the correction to $1$ inside the square bracket amounts to the
very small number $-0.017$ (inserting the central values of $\bar \ell_j$). 
Chiral corrections (even at two-loop order \cite{cola}) affect the isospin-two  
$\pi\pi$-interaction only very weakly and this feature seems to be reflected by
$\sigma_{\rm tot}(s)$ in Fig.\,12. Note however, that the argument made here is only 
suggestive and not rigorous, because the on-shell $\pi^-\pi^-\to \pi^-\pi^-$ final 
state interaction
does not factor out of the production amplitudes $A_1$ and $A_2$ in an obvious 
way. The same caveat applies to the $\pi^+\pi^-\to \pi^0\pi^0$ final state 
interaction which has been used as an argument for the observed enhancement in 
subsection 2.2.

Let us also comment on the theoretical uncertainties which are induced by the
present errorbars $\delta\bar \ell_j$ of the low-energy constants $\bar\ell_j$.
Taking again the total cross section at $\sqrt{s} =6m_\pi$ as a measure one finds
relative uncertainties of: $\pm 4.8\%$ from $\delta\bar \ell_1$,  $\pm 1.6\%$ 
from $\delta\bar \ell_2$, $\pm 0.3\%$ from $\delta\bar \ell_3$, and $\pm
1.0\%$ from $\delta\bar \ell_4$. As for the reaction $\pi^- \gamma\to\pi^-\pi^0
\pi^0$ the largest uncertainty goes along with $\bar \ell_1$ and adding them in
quadrature one gets a total relative uncertainty of $\pm 5.2\%$. This amounts 
again to a fairly accurate prediction. 

\begin{figure}
\begin{center}
\includegraphics[scale=.47,clip]{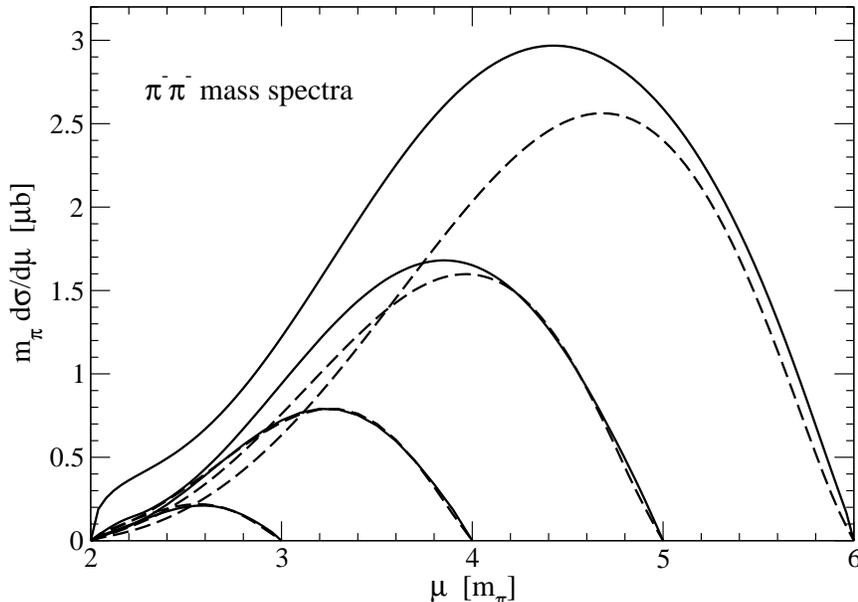}
\end{center}
\vspace{-1.cm}
\caption{$\pi^-\pi^-$ mass spectrum for the reaction $\pi^- \gamma\to  \pi^+\pi^-
\pi^-$ as a function of the $\pi^-\pi^-$ invariant mass $\mu$. The curves 
correspond to center-of-mass energies $\sqrt{s}=(4,5,6,7)m_\pi$ in ascending order.}
\end{figure}

The more exclusive observables than the total cross section are the two-pion 
mass spectra. Fig.\,13 shows the calculated $\pi^-\pi^-$ mass spectrum $m_\pi\, 
d\sigma/d\mu$ as a function of the $\pi^-\pi^-$ invariant mass $\mu$. The four 
pairs of (full and dashed) curves correspond to sections at 
center-of-mass-energies $\sqrt{s} =(4,5,6,7)m_\pi$ in ascending order. In 
essence Fig.13 reproduces the features of the total cross section, namely a
slight enhancement above $\sqrt{s}=6m_\pi$ by the next-to-leading order chiral 
corrections. The $\pi^+\pi^-$ mass spectrum $m_\pi\, d\sigma/d\sqrt{s_1}$
shown in Fig.\,14 indicates some more interesting structures. The dip of the
$\pi^+\pi^-$ mass spectrum at intermediate $\pi^+\pi^-$ invariant masses 
$\sqrt{s_1}$ becomes much more pronounced when including the next-to-leading
order chiral corrections. This distinctive feature holds e.g. at the 
center-of-mass energy $\sqrt{s} = 5m_\pi$ where the total cross sections in 
tree and one-loop approximation are equal. The $\pi^+\pi^-$ mass spectrum
of the reaction  $\pi^- \gamma\to\pi^+\pi^-\pi^-$ therefore seems to provide an 
interesting indicator for the role of chiral (pion-loop) dynamics beyond 
leading order. It is expected that the upcoming high-statistics data of the 
COMPASS experiment at CERN can reveal such dynamical details. Of course, the
squared (transversal) T-matrix $|\hat k \times (\vec q_1 A_1+\vec q_2 A_2)|^2$ 
with its full dependence on pion energies and angles incorporates still much 
more dynamical information.

\begin{figure}
\begin{center}
\includegraphics[scale=.47,clip]{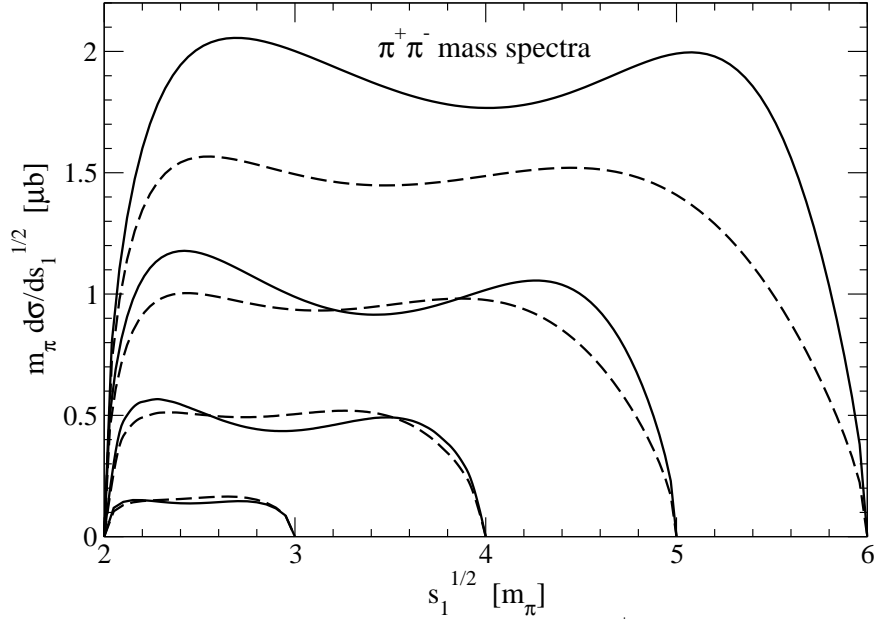}
\end{center}
\vspace{-1.cm}
\caption{$\pi^+\pi^-$ mass spectrum for the reaction $\pi^- \gamma \to\pi^+\pi^-
\pi^-$ as a function of the $\pi^+\pi^-$ invariant mass $\sqrt{s_1}$. The curves 
correspond to center-of-mass energies $\sqrt{s}=(4,5,6,7)m_\pi$ in ascending order.}
\end{figure}

In passing we note that $\sqrt{s}=(6-7)m_\pi$ is presumably the maximal 
center-of-mass energy up to which a one-loop calculation of the processes 
$\pi^- \gamma \to 3\pi$ in chiral perturbation theory can be trusted. At still
higher energies the contributions from meson resonances (such as $a_1(1260)$, 
$a_2(1320)$ etc.) will start to play a prominent role. In the context of such 
considerations it  should be kept in mind that the effects of the resonance tails
at low-energies are encoded in the empirical values of low-energy constants $\bar 
\ell_j$. The role of the low-lying $\rho(770)$ resonance occurring in the 
isospin-one $2\pi$-subsystem needs to be investigated by studying an appropriate 
resonance model for $\pi^- \gamma \to 3\pi$. Respecting fully gauge-invariance in
the construction of such a resonance model (with inclusion of finite resonance 
widths) represents some challenge. 
\vspace{-0.3cm}
\subsection*{Acknowledgments}
\vspace{-0.1cm}
I thank Jan Friedrich for many informative discussions. 

\end{document}